\documentclass{eptcs-modified}

\usepackage{amsmath}
\usepackage{amssymb}
\usepackage{url}
\usepackage{amsthm}
\usepackage{graphicx}

\newcommand{\ppad}{\textrm{PPAD}}
\newcommand{\fnp}{\textrm{FNP}}
\newcommand{\fp}{\textrm{FP}}

\begin{document}

\theoremstyle{definition}
\newtheorem{theorem}{Theorem}
\newtheorem{definition}[theorem]{Definition}
\newtheorem{problem}[theorem]{Problem}
\newtheorem{assumption}[theorem]{Assumption}
\newtheorem{corollary}[theorem]{Corollary}
\newtheorem{proposition}[theorem]{Proposition}
\newtheorem{lemma}[theorem]{Lemma}
\newtheorem{observation}[theorem]{Observation}
\newtheorem{fact}[theorem]{Fact}
\newtheorem{question}[theorem]{Open Question}
\newtheorem{conjecture}[theorem]{Conjecture}

\title{Efficient Decomposition of Bimatrix Games}


\author{
Xiang Jiang \& Arno Pauly
\institute{Computer Laboratory\\ University of Cambridge, United Kingdom}
\email{Arno.Pauly@cl.cam.ac.uk}
}
\def\titlerunning{Decomposing bimatrix games}
\def\authorrunning{X. Jiang \& A. Pauly}
\maketitle

\begin{abstract}
Exploiting the algebraic structure of the set of bimatrix games, a divide-and-conquer algorithm for finding Nash equilibria is proposed. The algorithm is fixed-parameter tractable with the size of the largest irreducible component of a game as parameter. An implementation of the algorithm is shown to yield a significant performance increase on inputs with small parameters.
\end{abstract}



\section{Introduction}
A bimatrix game is given by two matrices $(A, B)$ of identical dimensions. The first player picks a row $i$, the second player independently picks a column $j$. As a consequence, the first player receives the payoff $A_{ij}$, the second player $B_{ij}$. Both player are allowed to randomize over their choices, and will strive to maximize their expected payoff. A Nash equilibrium is a pair of strategies, such that no player can improve her expected payoff by deviating unilaterally.

If the payoff matrices are given by natural numbers, then there always is a Nash equilibrium using only rational probabilities. The computational task to find a Nash equilibrium of a bimatrix game is complete for the complexity class $\ppad$ \cite{papadimitrioub, daskalakis, dengd}. $\ppad$ is contained in $\fnp$, and commonly believed to exceed $\fp$. In particular, it is deemed unlikely that a polynomial-time algorithm for finding Nash equilibria exists.

The next-best algorithmic result to hope for could be a fixed-parameter tractable (fpt) algorithm \cite{downeyfellows, flum}, that is an algorithm running in time $f(k)p(n)$ where $n$ is the size of the game, $p$ a polynomial and $k$ a parameter. For such an algorithm to be useful, the assumption the parameter were usually small needs to be sustainable. The existence of fpt algorithms for finding Nash equilibria with various choices of parameters has been studied in \cite{estivill, hermelin, estivill2}.

In the present paper we demonstrate how \emph{products} and \emph{sums} of games -- and their inverse operations -- can be used to obtain a divide-and-conquer algorithm to find Nash equilibria. This algorithm is fpt, if the size of the largest component not further dividable is chosen as a parameter. \emph{Products} of games were introduced in \cite{paulyincomputabilitynashequilibria} as a means to classify the Weihrauch-degree \cite{brattka3, brattka2, paulybrattka, paulykojiro} of finding Nash equilibria for real-valued payoff matrices. \emph{Sums} appear originally in the PhD thesis \cite{paulyphd} of the second author; the algorithm we discuss was implemented in the Bachelor's thesis \cite{jiang} of the first author.

\section{Products and Sums of Games}
Both products and sums admit an intuitive explanation: The product of two games corresponds to playing both games at the same time, while the sum involves playing \emph{matching pennies} to determine which game to play, with one player being rewarded and the other one punished in the case of a failure to agree.

\subsection{Products}
In our definition of products, we let $[ \ , \ ] : \{1, \ldots, n\} \times \{1, \ldots, m\} \to \{1, \ldots, nm\}$ denote the usual bijection $[i, j] = (i - 1)n + j$. The relevant values of $n, m$ will be clear from the context. We point out that $[ \ , \ ]$ is polynomial-time computable and polynomial-time invertible.

\begin{definition}
Given an $n_1 \times m_1$ bimatrix game $(A^1, B^1)$ and an $n_2 \times m_2$ bimatrix game $(A^2, B^2)$, we define the $(n_1n_2) \times (m_1m_2)$ product game $(A^1, B^1) \times (A^2, B^2)$ as $(A, B)$ with $A_{[i_1, i_2][j_1, j_2]} = A_{i_1j_1}^1 + A_{i_2j_2}^2$ and $B_{[i_1, i_2][j_1, j_2]} = B_{i_1j_1}^1 + B_{i_2j_2}^2 $.
\end{definition}

\begin{theorem}
\label{bimatrix:theo:nashproduct1}
If $(x^k, y^k)$ is a Nash equilibrium of $(A^k, B^k)$ for both $k \in \{0, 1\}$, then $(x, y)$ is a Nash equilibrium of $(A, B)$, where $x_{[i_1i_2]} = x_{i_1}^1x_{i_2}^2$ and $y_{[m_1m_2]} = y_{m_1}^1y_{m_2}^2$.
\begin{proof}
We will prove that $x$ is a best response to $y$, if $x^k$ is a best response to $y^k$ for both $k \in \{0, 1\}$, the remaining part is analogous. By applying the following equivalence transformation
$$\begin{array}{cl} & \sum \limits_{i = 1}^{(n_1n_2)} \sum \limits_{j = 1}^{(m_1m_2)} x_i A_{i, j} y_j
\\ = & \sum \limits_{i_1 = 1}^{n_1} \sum \limits_{i_2 = 1}^{n_2} \sum \limits_{j_1 = 1}^{m_1} \sum \limits_{j_2 = 1}^{m_2} x_{[i_1, i_2]} A_{[i_1, i_2], [j_1, j_2]} y_{[i_1, j_2]}
\\ = & \sum \limits_{i_1 = 1}^{n_1} \sum \limits_{i_2 = 1}^{n_2} \sum \limits_{j_1 = 1}^{m_1} \sum \limits_{j_2 = 1}^{m_2} x^1_{i_1}x^2_{i_2} (A^1_{i_1, j_1} + A^2_{i_2, j_2}) y^1_{j_1} y^2_{j_2}
\\= & \left [ \sum \limits_{i_1 = 1}^{n_1} \sum \limits_{j_1 = 1}^{m_1} x_{i_1} A^1_{i_1, j_1} y_{j_1} \left (\sum \limits_{i_2 = 1}^{n_2} x_{i_2} \right ) \left ( \sum \limits_{j_2 = 1}^{m_2} y_{j_2} \right ) \right ] \\ & + \left [ \sum \limits_{i_2 = 1}^{n_2} \sum \limits_{j_2 = 1}^{m_2} x_{i_2} A^2_{i_2, j_2} y_{j_2} \left (\sum \limits_{i_1 = 1}^{n_1} x_{i_1} \right ) \left ( \sum \limits_{j_1 = 1}^{m_1} y_{j_1} \right ) \right ]
\\= & \left [ \sum \limits_{i_1 = 1}^{n_1} \sum \limits_{j_1 = 1}^{m_1} x_{i_1} A^1_{i_1, j_1} y_{j_1} \right] + \left [ \sum \limits_{i_2 = 1}^{n_2} \sum \limits_{j_2 = 1}^{m_2} x_{i_2} A^2_{i_2, j_2} y_{j_2} \right ] \end{array}$$
on both sides of the best response condition $$\sum \limits_{i = 1}^{n_1n_2} \sum \limits_{j = 1}^{m_1m_2} x_i A_{i, j} y_j \geq \sum \limits_{i = 1}^{n_1n_2} \sum \limits_{j = 1}^{m_1m_2} \hat{x}_i A_{i, j} y_j$$ one obtains the following form for the best response condition:
$$\begin{array}{rcl} \left [ \sum \limits_{i_1 = 1}^{n_1} \sum \limits_{j_1 = 1}^{m_1} x_{i_1} A^1_{i_1, j_1} y_{j_1} \right] & + & \left [ \sum \limits_{i_2 = 1}^{n_2} \sum \limits_{j_2 = 1}^{m_2} x_{i_2} A^2_{i_2, j_2} y_{j_2} \right ] \\ & \geq & \\ \left [ \sum \limits_{i_1 = 1}^{n_1} \sum \limits_{j_1 = 1}^{m_1} \hat{x}_{i_1} A^1_{i_1, j_1} y_{j_1} \right] & + & \left [ \sum \limits_{i_2 = 1}^{n_2} \sum \limits_{j_2 = 1}^{m_2} \hat{x}_{i_2} A^2_{i_2, j_2} y_{j_2} \right ]\end{array}$$ As this is just the sum of the best response conditions for the individual games $(A^1, B^2)$ and $(A^2, B^2)$, the claim follows.
\end{proof}
\end{theorem}

\begin{theorem}
\label{bimatrix:theo:nashproduct2}
If $(x, y)$ is a Nash equilibrium of $(A, B)$, then $(x^1, y^1)$ given by $x_{i}^1 = \sum \limits_{l = 1}^{n_2} x_{[i,l]}$ and $y_{j}^1 = \sum \limits_{l = 1}^{m_2} y_{[j,l]}$ is a Nash equilibrium of $(A^1, B^1)$.
\begin{proof}
Again the proof uses contraposition. Assume w.l.o.g. that $\hat{x}^1$ is a better response against $y^1$ than $x^1$, that is: $$\sum \limits_{i = 1}^{n_1} \sum \limits_{p = 1}^{m_1} \hat{x}^1_i A^1_{i,j} y^1_j >  \sum \limits_{i = 1}^{n_1} \sum \limits_{j = 1}^{m_1} x^1_i A^1_{i,j} y^1_j$$ Add $\sum \limits_{i = 1}^{n_2} \sum \limits_{j = 1}^{m_2} x^2_i A^2_{i,j} y^2_j$ on both sides, and apply the reverse of the transformation used in the proof of Theorem \ref{bimatrix:theo:nashproduct1}. Then one obtains: $$\sum \limits_{i = 1}^{n_1n_2} \sum \limits_{j = 1}^{m_1m_2} \hat{x}_i A_{i, j} y_j > \sum \limits_{i = 1}^{n_1n_2} \sum \limits_{j = 1}^{m_1m_2} \hat{x}_i A_{i, j} y_j$$ with $\hat{x}$ defined via $\hat{x}_{[i_1, i_2]} = \hat{x}^1_{i_1}x^2_{i_2}$. This contradicts the assumption that $x$ would be a best response against $y$, so $(x, y)$ cannot be a Nash equilibrium.
\end{proof}
\end{theorem}

\subsection{Sums}
The sum of games involves another parameter besides the two component games, which just is a number exceeding the absolute value of all the payoffs.
\begin{definition}
Given an $n_1 \times m_1$ bimatrix game $(A^1, B^1)$ and an $n_2 \times m_2$ bimatrix game $(A^2, B^2)$, we define the $(n_1 + n_2) \times (m_1 + m_2)$ sum game $(A^1, B^1) + (A^2, B^2)$  via the constant $K > \max_{i,j} \{|A_{i,j}|, B_{i,j}|\}$ as $(A, B)$ with: $$A_{i,j} = \begin{cases} A_{ij}^1 & \textnormal{if } i \leq n_1, j \leq m_1 \\ A_{(i-n_1),(j-m_1)}^2 & \textnormal{if } i > n_1, j > m_1 \\ K & \textnormal{otherwise} \end{cases}$$ $$B_{i,j} = \begin{cases} B_{ij}^1 & \textnormal{if } i \leq n_1, j \leq m_1 \\ B_{(i-n_1),(j-m_1)}^2 & \textnormal{if } i > n_1, j > m_1 \\ - K & \textnormal{otherwise} \end{cases}$$
\end{definition}

\begin{lemma}
\label{bimatrix:lemma:summatchingpennies}
Let $(x, y)$ be a Nash equilibrium of $(A^1, B^1) + (A^2, B^2)$. Then $0 < \left ( \sum_{i = 1}^{n_1} x_i \right ) < 1$ and $0 < \left ( \sum_{j = 1}^{m_1} y_j \right ) < 1$.
\begin{proof}
Assume $(x, y)$ is a Nash equilibrium. The following circular reasoning demonstrates that any of the forbidden cases yields a contradiction.
\begin{enumerate}
\item  If $0 = \sum_{i = 1}^{n_1} x_i$, then also $0 = \sum_{j = 1}^{m_1} y_j$.

If $y_j > 0$ for any $j \leq m_1$, then $y'$ defined via $y'_j = 0$, $y'_{m_1 + 1} = y_{m_1 + 1} + y_j$ and $y'_l = y_l$ for $l \neq j, m_1 + 1$ is a better response against $x$ than $y$: The payoff difference between $y'$ and $y$ is $y_j \left (\sum_{i_2 = 1}^{n_2} (B_{i_2,m_1 + 1}^2 + K)x_{n_1 + i_2}\right )$, and by choice of $K$ every $(B_{i_2,m_1 + 1}^2 + K)$ is positive.

\item If $0 = \sum_{j = 1}^{m_1} y_j$, then $\sum_{i = 1}^{n_1} x_i = 1$.

If $x_i > 0$ for any $i > n_1$, then $x'$ defined via $x'_i = 0$, $x'_1 = x_1 + x_i$ and $x'_l = x_l$ for $l \neq 1, i$ is a better response against $y$ than $x$: The payoff difference between $x'$ and $y$ is $x_i \left (\sum_{j_2 = 1}^{m_2} (- A_{i - n_1,j_2}^2 + K)y_{m_1 + j_2}\right )$, and by choice of $K$ every $(- A_{i - n_1,j_2}^2 + K)$ is positive.

\item If $\sum_{i = 1}^{n_1} x_i = 1$, then $\sum_{j = 1}^{m_1} y_j = 1$.

The proof proceeds as in 1. via symmetry.

\item If $\sum_{j = 1}^{m_1} y_j = 1$, then $0 = \sum_{i = 1}^{n_1} x_i$.

The proof proceeds as in 2. via symmetry.
\end{enumerate}
\end{proof}
\end{lemma}

\begin{theorem}
\label{bimatrix:theo:nashsum2}
If $(x, y)$ is a Nash equilibrium of $(A^1, B^1) + (A^2, B^2)$, then a Nash equilibrium $(x^1, y^1)$ of $(A^1, B^1)$ can be obtained as $x^1_i = \frac{x_i}{\sum_{l = 1}^{n_1} x_l}$ and $y^1_j = \frac{y_i}{\sum_{l = 1}^{m_1} y_l}$.
\begin{proof}
By Lemma \ref{bimatrix:lemma:summatchingpennies}, $(x^1, y^1)$ is well-defined, and clearly a strategy profile. W.l.o.g. we assume that $x^1$ is not a best response against $y^1$, and derive a contradiction. Let $\overline{x}^1$ be a better response against $y^1$ than $x^1$. Define $\overline{x}$ via $\overline{x}_i = \overline{x}^1_i \left ( \sum_{l = 1}^{n_1} x_l \right )$ for $i \leq n_1$, and $\overline{x}_i = x_i$ otherwise. We claim that $\overline{x}$ is a better response against $y$ than $x$. The payoff difference between $\overline{x}$ and $x$ for the first player can readily be computed to be $\left ( \sum_{l = 1}^{n_1} x_l \right ) \sum_{i_1 = 1}^{n_1} \sum_{j_1 = 1}^{m_1} A_{i_1,j_1} (\overline{x}^1_{i_1} - x^1_{i_1})y_{i_2}$. Up to the positive factor $\left [\left ( \sum_{l = 1}^{n_1} x_l \right )\left ( \sum_{k = 1}^{m_1} y_k \right )\right ]^{-1}$, this is equal to the payoff difference between $\overline{x}^1$ and $x^1$, hence the former is positive iff the latter is.
\end{proof}
\end{theorem}

\begin{theorem}
\label{bimatrix:theo:nashsum1}
Let $(x^k, y^k)$ be a Nash equilibrium of $(A^k, B^k)$ resulting in payoffs $(P^k, Q^k)$ for both $k \in \{1, 2\}$. Then $(x, y)$ is a Nash equilibrium of $(A^1, B^1) + (A^2, B^2)$, where $x_i = x^1_i \frac{K - Q^2}{2K - Q^1 - Q^2}$ for $i \leq n_1$, $x_i = x^2_{i - n_1} \frac{K - Q^1}{2K - Q^1 - Q^2}$ for $i > n_1$, $y_j = y^1_j \frac{K - P^2}{2K - P^1 - P^2}$ for $j \leq m_1$, $y_j = y^2_{j - m_1} \frac{K - P^1}{2K - P^1 - P^2}$ for $j > m_1$.
\begin{proof}
In the given situation, assume $(x, y)$ were not a Nash equilibrium. W.l.o.g., let this be due to $x$ not being a best response to $y$. This is equivalent to the existence of some $i$ with $x_i > 0$, but the pure strategy $i$ is not a best response to $y$. The latter means there is a better response $k$ to $y$, i.e.:
\begin{description}
\item[Case $i \leq n_1, k \leq n_1$] {\footnotesize \begin{align*} & \left ( \sum_{j_1 = 1}^{m_1} A_{kj_1}^1y^1_{j_1}\frac{K - P^2}{2K - P^1 - P^2}\right ) + \left ( \sum_{j_2 = 1}^{m_2} Ky^2_{j_2}  \frac{K - P^1}{2K - P^1 - P^2} \right )\\ > & \left ( \sum_{j_1 = 1}^{m_1} A_{ij_1}^1y^1_{j_1}\frac{K - P^2}{2K - P^1 - P^2}\right ) + \left ( \sum_{j_2 = 1}^{m_2} Ky^2_{j_2}  \frac{K - P^1}{2K - P^1 - P^2} \right )\end{align*}}
    Subtracting {\footnotesize $ \left ( \sum_{j_2 = 1}^{m_2} Ky^2_{j_2}  \frac{K - P^1}{2K - P^1 - P^2} \right )$} on both sides, then dividing by {\footnotesize $\frac{K - P^2}{2K - P^1 - P^2}$} shows that $k$ is a better response against $y^1$ than $i$ in the game $(A^1, B^1)$. But then $x^1_i = 0$ follows, hence $x_i = 0$ in contradiction to the assumption.
\item[Case $i \leq n_1, k > n_1$] {\footnotesize \begin{align*} & \left ( \sum_{j_1 = 1}^{m_1} K y^1_{j_1}\frac{K - P^2}{2K - P^1 - P^2}\right ) + \left ( \sum_{j_2 = 1}^{m_2} A^1_{kj_2}y^2_{j_2}  \frac{K - P^1}{2K - P^1 - P^2} \right )\\ > & \left ( \sum_{j_1 = 1}^{m_1} A_{ij_1}^1y^1_{j_1}\frac{K - P^2}{2K - P^1 - P^2}\right ) + \left ( \sum_{j_2 = 1}^{m_2} Ky^2_{j_2}  \frac{K - P^1}{2K - P^1 - P^2} \right )\end{align*}}
    As they are sums over stochastic vectors, we find $\left (\sum_{j_1}^{m_1} y^1_{j_1} \right ) = \left (\sum_{j_2}^{m_2} y^2_{j_2} \right ) = 1$. Moreover, $x_i \neq 0$ implies $x^1_i \neq 0$, and this in turn implies $ \left ( \sum_{j_1 = 1}^{m_1} A_{ij_1}^1y^1_{j_1} \right ) = P^1$. Hence, after multiplying both sides by $(2K - P^1 - P^2)$ the previous inequality simplifies to: {\footnotesize
    \[K(K - P^2) + \left ( \sum_{j_2 = 1}^{m_2} A^1_{kj_2}y^2_{j_2} \right )(K - P^1) >  P^1(K - P^2) + K (K - P^1)\]}
    This in turn can be simplified to $\left ( \sum_{j_2 = 1}^{m_2} A^1_{kj_2}y^2_{j_2} \right ) > P^2$, which contradicts the assumption $P^2$ were the optimal payoff achievable by player $1$ against $y^2$.
\item[Case $i > n_1, k \leq n_1$] Analogous to {\bf Case} $i \leq n_1, k > n_1$
\item[Case $i > n_1, k > n_1$] Analogous to {\bf Case} $i \leq n_1, k \leq n_1$
\end{description}
\end{proof}
\end{theorem}

\section{The algorithm}
\label{sec:algo}
Our basic algorithm proceeds as follows: To solve a game $(A, B)$
\begin{enumerate}
\item test whether $(A, B)$ is the sum of $(A^1, B^1)$ and $(A^2, B^2)$ via some constant $K$. If yes, solve $(A^1, B^1)$ and $(A^2, B^2)$ and combine their Nash equilibria to an equilibrium of $(A, B)$ via Theorem \ref{bimatrix:theo:nashsum1}. If no,
\item test whether $(A, B)$ is the product of $(A^1, B^1)$ and $(A^2, B^2)$. If yes, solve $(A^1, B^1)$ and $(A^2, B^2)$ and combine their Nash equilibria to an equilibrium of $(A, B)$ via Theorem \ref{bimatrix:theo:nashproduct1}. If no,
\item find a Nash equilibrium of $(A, B)$ by some other means.
\end{enumerate}

For some $n \times m$ game $(A, B)$ let let $S(A, B)$ denote its size, i.e.~$S(A, B) = nm$, and let $\lambda(A, B)$ be the size of the largest game for which $3.$ in our algorithm is called. Let $f(k)$ be the time needed for the external algorithm called in $3.$ on a game of size $k$. Then the runtime of our algorithm is bounded by $O(S^3f(\lambda))$, in particular, it is an $fpt$-algorithm:

Testing whether a game is a sum, and computing the components, if applicable, can be done in linear time. The sum of the sizes of the components is less than the size of the original game. Finally, combining Nash equilibria can be done in linear time, too.

Whether a game is a product of factors of a fixed size can also be tested in linear time. Testing the different possible factors yields quadratic time for this part. This already includes computing the components, and the product of the sizes of the factors is equal to size of the original game. Again, combining the Nash equilibria takes linear time.

As a slight modification of our algorithm, one can eliminate (iteratively) strictly dominated strategies at each stage of the algorithm. We recall that a strategy $i$ of some player is called strictly dominated by some other strategy $j$, if against any strategy chosen by the opponent, $i$ provides its player with a strictly better payoff than $j$. A strictly dominated strategy can never be used in a Nash equilibrium. It is easy to verify that a game decomposable as a sum never has any strictly dominated strategies, but may occur as the result of the elimination of such strategies. Hence, including an elimination step for each stage increasing the potential for decomposability. Elimination of strictly dominated strategies commutes with decomposition of products, i.e.~the reduced from of the product is the product of the reduced forms of the factors. The algorithm remains $fpt$ if such a step is included. A detailed investigation of complexity issues regarding removal of dominated strategies can be found in \cite{paulycomplexityofise}.

\section{Empirical evaluation}
Only a small fraction of the bimatrix games of a given size and bounded integer payoffs will be decomposable by our techniques, this limiting the applicability of the algorithm in Section \ref{sec:algo}. However, to some extent we can expect patterns in the definitions of real-world game situation to increase the decomposability of the derived bimatrix games. For example, the structure of Poker-style games implies decomposability, as can be concluded from the considerations in \cite{gilpin}.

To obtain a first impression whether using the decomposition algorithm is indeed beneficial, a collection of 100 random decomposable games was created. Each game has 95-105 strategies per player, and payoff values range from 0 to 50. The decomposability was ensure by creating a random tree representing the relevant decomposition structure first, using probabilities of 0.4 each for sum and product decomposition, and of 0.2 for an elimination of strictly dominated strategies step. The height of the trees was limited to 80, additionally vertices corresponding to games of size up to 6 were turned into leaves. At the leaves, the payoffs were chosen uniformly subject to the constraints derived from the structure and the overall constraint of payoff values being between 0 and 50. Finally, the corresponding bimatrix games were computed.

Both as a benchmark, and in order to compute Nash equilibria of the irreducible component games, the tool \textsc{Gambit} \cite{gambit} was used. \textsc{Gambit} offers a variety of algorithm for computing Nash equilibria of bimatrix games, we used:
\begin{enumerate}
\item gambit-enummixed: using extreme point enumeration
\item gambit-gnm: using a global Newton method approach
\item gambit-lcp: using linear complementarity
\item gambit-simpdiv: using simplicial subdivision
\end{enumerate}

\begin{figure}[htbp]
\centering
\includegraphics[width=\textwidth]{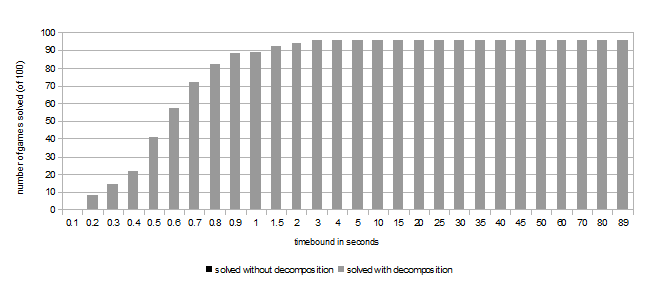}
\caption{gambit-enummixed}
\label{enum}
\end{figure}

\begin{figure}[htbp]
\centering
\includegraphics[width=\textwidth]{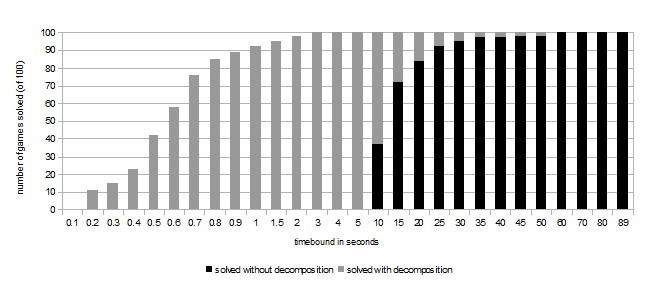}
\caption{gambit-gnm}
\label{enum}
\end{figure}

\begin{figure}[htbp]
\centering
\includegraphics[width=\textwidth]{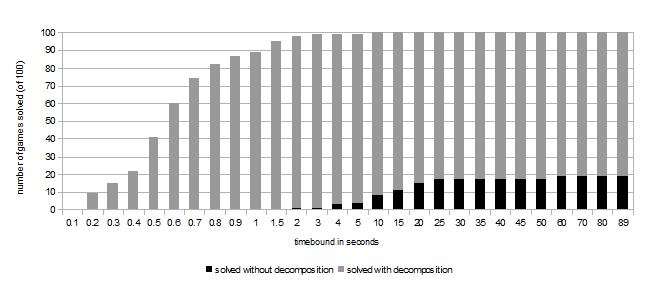}
\caption{gambit-lcp}
\label{enum}
\end{figure}

\begin{figure}[htbp]
\centering
\includegraphics[width=\textwidth]{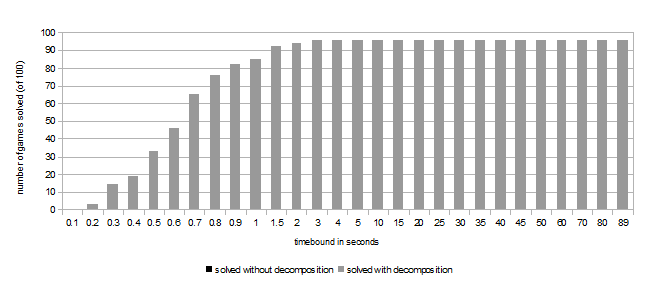}
\caption{gambit-simpdiv}
\label{enum}
\end{figure}

Figures 1.-4. show for each of the \textsc{Gambit} algorithms how many of our decomposable example games could be solved in some given time bound (per game, not total) using only the \textsc{Gambit} algorithm directly, or exploiting decomposition implemented in C++ first. Despite the fact that our decomposition algorithm was not optimized, it turned out that using decomposition almost all games could be solved in under 3 seconds, whereas even gambit-gnm as the fastest \textsc{Gambit} algorithm on the sample took 30 seconds for a similar feat. Thus, there is clear indication that on suitable data, exploiting the algebraic structure underlying the decomposition algorithm yields a significant increase in performance.
\bibliographystyle{eptcs}
\bibliography{../spieltheorie}

\end{document}